\documentclass[acmsmall]{acmart}

\AtBeginDocument{%
  }

\setcopyright{acmlicensed}
\acmDOI{10.1145/3715779}
\acmYear{2025}
\acmJournal{PACMSE}
\acmVolume{2}
\acmNumber{FSE}
\acmArticle{FSE060}
\acmMonth{7}
\received{2024-09-13}
\received[accepted]{2025-01-14}

\usepackage{tabularray}
\usepackage{array}
\usepackage{pifont}
\usepackage{tcolorbox}
\usepackage{rotating}
\usepackage{enumitem}
\usepackage{makecell}
\usepackage{array}
\usepackage{graphicx}
\usepackage{multirow}
\usepackage{ragged2e}
\usepackage{longtable}
\usepackage{geometry}
\usepackage{listings}
\usepackage{tabularx}
\usepackage{subcaption}
\usepackage{booktabs}
\usepackage{todonotes}
\usepackage{soul}

\newenvironment{mybox}{%
    \begin{tcolorbox}[title={Summary}]%
    }{ 
    \end{tcolorbox}
}

\begin{document}

\title{Protecting Privacy in Software Logs: What Should Be Anonymized?}

\author{Roozbeh Aghili}
\email{roozbeh.aghili@polymtl.ca}
\affiliation{%
  \institution{Polytechnique Montréal}
  \city{Montréal}
  \country{Canada}}
\author{Heng Li}
\authornote{Corresponding author.}
\email{heng.li@polymtl.ca}
\affiliation{%
  \institution{Polytechnique Montréal}
  \city{Montréal}
  \country{Canada}}
\author{Foutse Khomh}
\email{foutse.khomh@polymtl.ca}
\affiliation{%
  \institution{Polytechnique Montréal}
  \city{Montréal}
  \country{Canada}}

\begin{abstract}
\label{sec:abstract}
Software logs, generated during the runtime of software systems, are essential for various development and analysis activities, such as anomaly detection and failure diagnosis. However, the presence of sensitive information in these logs poses significant privacy concerns, particularly regarding \textit{Personally Identifiable Information (PII)} and quasi-identifiers that could lead to re-identification risks. While general data privacy has been extensively studied, the specific domain of privacy in software logs remains underexplored, with inconsistent definitions of sensitivity and a lack of standardized guidelines for anonymization. To mitigate this gap, this study offers a comprehensive analysis of privacy in software logs from multiple perspectives. We start by performing an analysis of 25 publicly available log datasets to identify potentially sensitive attributes. Based on the result of this step, we focus on three perspectives: privacy regulations, research literature, and industry practices. We first analyze key data privacy regulations, such as the \textit{General Data Protection Regulation (GDPR)} and the \textit{California Consumer Privacy Act (CCPA)}, to understand the legal requirements concerning sensitive information in logs. Second, we conduct a systematic literature review to identify common privacy attributes and practices in log anonymization, revealing gaps in existing approaches. Finally, we survey 45 industry professionals to capture practical insights on log anonymization practices. Our findings shed light on various perspectives of log privacy and reveal industry challenges, such as technical and efficiency issues while highlighting the need for standardized guidelines. By combining insights from regulatory, academic, and industry perspectives, our study aims to provide a clearer framework for identifying and protecting sensitive information in software logs.
\end{abstract}

\ccsdesc[500]{Software and its engineering~Software safety}
\ccsdesc[500]{Security and privacy~Privacy protections}

\keywords{Software logs, Data privacy, Anonymization, Log sensitivity, GDPR}

\maketitle

\section{Introduction}
\label{sec:introduction}

Software logs are record statements used by software developers to capture valuable runtime information about software systems. They are generated by logging statements inserted into the source code, which produce execution logs during runtime~\cite{barik2016bones, li2020qualitative}. These logs are then either appended to log files for later analysis or output to the standard console for immediate monitoring. Logging is crucial for tasks such as failure diagnosis and anomaly detection~\cite{cinque2012event, batoun2024literature, qin2024preprocessing, aghili2023studying}. For instance, logs are often the only resource available for diagnosing field failures~\cite{yuan2010sherlog}. By recording different events and activities within a software system, logs help developers and other software practitioners manage and maintain the health and performance of their systems effectively.

As the number of log statements in software systems increases, the likelihood of capturing sensitive information within logs also rises, making privacy preservation more critical~\cite{jain2016big}. Protecting privacy is essential not only to comply with regulatory requirements but also to maintain user trust and prevent potential harm that could arise from data leaks. A significant concern in software log management and sharing is the exposure of \textit{Personally Identifiable Information (PII)}. PII refers to any data that can be used to identify an individual, either directly, such as names and email addresses, or indirectly, such as device identifiers. While PII is often the focus of privacy concerns, it is not the only source of re-identification of individuals. A study by Sweeney et al.~\cite{sweeney2002k} demonstrated that using just three attributes—postal code, date of birth, and sex—over 86\% of U.S. citizens could be uniquely identified. These attributes, known as \textit{quasi-identifiers}, may not reveal any information when used individually, but when combined, they can effectively re-identify individuals.

While data privacy is a well-defined area with several decades of research, the specific domain of privacy in software logs remains largely underexplored. Despite extensive work in general data privacy, there is a significant gap in our understanding of what constitutes sensitive information within software logs. We lack clear agreement on which log attributes should be considered sensitive and which are harmless. Moreover, there is little knowledge about which combinations of log attributes might lead to the re-identification of individuals or entities. Existing studies on log privacy and anonymization~\cite{gu2023pd, mcsherry2010differentially, xu2002prefix} often develop their own definitions of sensitivity, leading to inconsistent approaches and a fragmented understanding of the risks involved. This inconsistency highlights the need for a systematic exploration of log sensitivity to establish a clearer framework for protecting privacy in this context.

A significant challenge in the field of software log privacy is the reluctance of organizations to share their datasets, including software logs and traces. Privacy concerns largely drive this hesitation, as the risk of exposing sensitive information can be substantial. As a result, there is a notable lack of publicly available datasets for log analysis tasks such as anomaly detection~\cite{aghili2023studying, bogatinovski2021artificial}. Besides, there have been incidents upon publishing anonymized data where the public datasets led to the identification of users. For example, in 2006, AOL released a dataset containing 20 million keyword searches made by its users. Although the data was anonymized, the presence of PII led to identifying several individuals by analyzing the unique search queries associated with them~\cite{aol, cooper2008survey}. Similarly, the Netflix Prize dataset, which contained anonymous movie ratings of 500,000 subscribers, was found to be vulnerable to de-anonymization. Narayanan et al.~\cite{narayanan2006break} demonstrated that by cross-referencing the Netflix data with information from the \textit{Internet Movie Database (IMDb)}, they could identify individual subscribers, revealing their political preferences and other potentially sensitive information. In the context of software logs, similar risks exist as logs often contain detailed information about system events, user activities, and configuration settings. If such logs are anonymized inadequately, there is a risk of re-identifying individuals or revealing sensitive information, necessitating effective privacy-preserving methods in log management.

To address privacy challenges in software logs and better understand sensitive information, we conduct a comprehensive study. We examine the topic from multiple perspectives. First, we analyze 25 publicly available log datasets to identify common attributes in software logs. We then review key privacy regulations, such as \textit{General Data Protection Regulation (GDPR)} and \textit{California Consumer Privacy Act (CCPA)}, to understand legal requirements related to data privacy, framing privacy expectations and risks. Next, we review existing research and tools on log anonymization to identify common privacy attributes and gaps in the field. Finally, we survey industry professionals to gather insights on privacy concerns in software logs. By combining findings from log analysis, regulatory analysis, literature review, and industry perspectives, our study aims to provide a well-rounded understanding of which attributes in software logs should be anonymized to protect privacy effectively. We specifically aim to answer the following \textit{Research Questions (RQ)}.

\begin{itemize}
\item [RQ1] \textbf{\textit{What are the most common attributes in software logs?}}
In this first RQ, we explore the information collected in different software logs. By identifying the most common log attributes, we can better assess the potential sensitivity of the data being logged and ensure compliance with privacy regulations.

\item[RQ2] \textbf{\textit{What do privacy regulations define as sensitive information in software logs?}}
In this RQ, we inspect the regulations related to data privacy. By analyzing these regulations, we aim to clarify the legal obligations that organizations must adhere to when handling log data, ensuring that privacy is maintained.

\item[RQ3] \textbf{\textit{What do existing research and tools define as sensitive attributes in software logs?}}
Existing research and tools for log anonymization provide valuable insights into what the academic community considers sensitive in software logs. By exploring these definitions, we uncover common practices and gaps that need to be addressed to enhance log privacy.

\item[RQ4] \textbf{\textit{What do industry professionals consider sensitive in software logs?}}
Industry professionals often deal with the practical challenges of anonymizing software logs in real-world scenarios. By surveying these experts, we aim to capture the attributes they prioritize as sensitive and understand how they approach log privacy, providing a practical perspective that complements the regulatory and academic views.
\end{itemize}

Our work makes several contributions:

\begin{enumerate}
\item We provide the first in-depth analysis of privacy concerns in software logs, considering multiple perspectives, including regulatory frameworks, research, and industry practices.

\item We analyze 25 log datasets to understand commonly collected information in software logs.

\item Our \textit{Systematic Literature Review (SLR)} offers a comprehensive overview of the existing practices and tools in the domain of log privacy.

\item Through a survey of 45 industry professionals, we uncover practical insights into the real-world challenges and strategies involved in log anonymization.
\end{enumerate}

The rest of the paper is organized as follows. Section~\ref{sec:preliminary_study} analyzes the common attributes in software logs. Section~\ref{sec:regulations} explores data privacy regulations. Section~\ref{sec:articles_tools} reviews relevant literature and tools on log privacy. Section~\ref{sec:industry} presents insights from our industry survey. Section~\ref{sec:discussion} discusses the implications of our findings, while section~\ref{sec:threats} addresses threats to validity. Finally, Section~\ref{sec:conclusion} concludes the paper, and Section~\ref{sec:data_availability} provides our replication package for future studies.
\section{Common Attributes in Software Logs}
\label{sec:preliminary_study}

Software logs can have diverse forms and structures, varying widely in attributes, format, and content. Logs may range from simple text files recording basic events to complex formats capturing detailed system behaviors, user interactions, or security-related incidents. This variability reflects the different purposes and contexts of log generation, making it challenging to standardize or categorize them. Understanding what constitutes sensitive information within these logs becomes a complex issue, which is why in this section, we analyze a selection of publicly available logs and identify common attributes that could potentially be considered sensitive.

\subsection{Approach}
In our analysis, we examine 25 log datasets drawn from two well-known collections: LogHub~\cite{zhu2023loghub} and the \textit{Internet Traffic Archive (ITA)}~\cite{ita}. These datasets provide a wide range of logs from different domains and system types, enabling us to investigate various attributes and formats. Specifically, we analyze all datasets from the LogHub collection, which includes logs from \textit{distributed systems} (HDFS, Hadoop, Spark, Zookeeper, OpenStack), \textit{supercomputers} (BGL, HPC, Thunderbird), \textit{operating systems} (Windows, Linux, Mac), \textit{mobile systems} (Android, HealthApp), \textit{server applications} (Apache, OpenSSH), and \textit{standalone software} (Proxifier). Additionally, from the ITA collection, we focus on \textit{web application} logs, including datasets from Calgary, Saskatchewan, Boston, WorldCup98, NASA, ClarkNet, EPA, and SDSC. We also incorporate the YouTube dataset collected by Zink et al.~\cite{youtube}. For detailed information about each dataset, please refer to the references. 

To identify common attributes across different log datasets, we start by parsing each of the 25 log datasets using Drain~\cite{he2017drain}, a well-known log parsing tool. Specifically, we analyze a sample of 2,000 log lines per dataset to extract log templates. We identify up to 350 unique log templates per dataset and then examine a subset of example log lines for each template to determine potential attributes. These attributes correspond to variables in the log structure, which Drain marks with an asterisk (*). To assess potential sensitive attributes, we take an inclusive approach, collecting any attribute that could pose a vulnerability. We later refine this by considering different perspectives on sensitivity (see Sections~\ref{sec:regulations}, \ref{sec:articles_tools}, and \ref{sec:industry}) and ultimately defining our own criteria (see Section~\ref{sec:discussion}).

Table~\ref{tab:log_examples} presents four sample logs from ClarkNet, OpenSSH, Hadoop, and HPC datasets. As can be seen, each log follows a distinct pattern, provides specific information, and includes particular attributes. The first log is a typical HTTP access log entry showing a request made to a web server. This log has attributes such as IP address, request method, and status code. The second log indicates a failed authentication attempt via SSH and contains details such as date and time, host information, and process ID. The third log shows the maximum resource capabilities for containers in Hadoop MapReduce, with attributes such as log level and configuration details. Finally, the fourth log sample reports the temperature of a node, providing attributes such as node ID and environmental data.

\begin{table}[t]
\caption{Example software logs from different datasets.}
\footnotesize
\centering
\begin{tabularx}{\textwidth}{|>{\raggedright\arraybackslash}X|}
\hline
\textbf{1.} \texttt{132.201.126.207 - - [28/Aug/1995:17:23:48 -0400] "GET /pub/atomicbk/images/spicy2.gif HTTP/1.0" 200 29018} \\
\hline
\textbf{2.} \texttt{Dec 10 07:07:38 LabSZ sshd[24206]: pam\_unix(sshd:auth): authentication failure; logname= uid=0 euid=0 tty=ssh ruser= rhost=ec2-52-80-34-196.cn-north-1.compute.amazonaws.com.cn} \\
\hline
\textbf{3.} \texttt{2015-10-18 18:01:53,713 INFO [main] org.apache.hadoop.mapreduce.v2.app.rm.RMContainerAllocator: maxContainerCapability: <memory:8192, vCores:32>} \\
\hline
\textbf{4.} \texttt{76723 node-55 node temperature 1077205904 ambient=33} \\
\hline
\end{tabularx}
\label{tab:log_examples}
\end{table}

\subsection{Results}
Through analysis of 25 log datasets, \textbf{we identify 18 frequently shared attributes in software logs}. Detailed in Table~\ref{tab:attributes_preliminary}, the table includes definitions, examples, and frequency, which represents the percentage of log datasets containing each attribute. \textbf{The most common attributes are timestamps (100\%), IP addresses (80\%), file paths (72\%), IDs (72\%), and components (60\%)}. 

\begin{table}[t]
\centering
\caption{The most frequent log attributes and examples}
\footnotesize 
\resizebox{\textwidth}{!}{%
\begin{tabular} {|p{0.02\textwidth} |p{0.18\textwidth} |p{0.34\textwidth} |p{0.42\textwidth} |p{0.04\textwidth}|}
\hline
\textbf{ID} & \textbf{Attribute} & \textbf{Definition} & \textbf{Example} & \textbf{Freq. (\%)} \\ \hline
1  & Timestamp           & Date and time of the log entry. & 2024-08-15 - 12:11:37 & 100 \\ \hline
2  & IP address              & Unique number for network devices. & 192.168.1.1 & 80 \\ \hline
3  & File path               & Location of a file in the filesystem. & /user/root/rand/\_temporary/part-00742 & 72 \\ \hline
4  & IDs                     & Identifiers for system entities. & Process ID, Thread ID, Job ID, Node ID, Application ID, Device ID & 72 \\ \hline
5  & Component               & Module of the system generating the log. & org.apache.hadoop.mapreduce.v2.app.MRAppMaster & 60 \\ \hline
6  & Hostname & Unique name for network devices.  & ec2-52-80-34-196.cn-north-1.compute.amazonaws.com.cn & 44 \\ \hline
7  & Log level               & Severity of the log event. & INFO & 40 \\ \hline
8 & Port number             & Number identifying a specific service. & 8080 & 36 \\ \hline
9 & Request protocol        & Protocol used for the request. & HTTP/1.0 & 36 \\ \hline
10 & Request status code     & HTTP status code returned by the server. & 200 & 36 \\ \hline
11 & Request response size   & Size of the server's response. & 56 B & 36 \\ \hline
12 & Configuration details   & System configuration information. & vCores:32 & 36 \\ \hline
13 & Request method          & Method used to request a resource. & GET & 32 \\ \hline
14  & URL                     & Address of resources on the internet. & http://cs-www.bu.edu/lib/pics/bu-logo.gif & 24 \\ \hline
15  & MAC address             & Unique identifier for network interfaces. & 5c:50:15:4c:18:13 & 8 \\ \hline
16 & Request response time   & Time taken for server response. & 0.3 s & 8 \\ \hline
17 & Environmental data      & Data related to environmental conditions. & temperature ambient=33 & 8 \\ \hline
18 & Username   & Unique user identifier. & cheng & 8 \\ \hline
\end{tabular}%
}
\label{tab:attributes_preliminary}
\end{table}

An IP address is often the first attribute identified as sensitive in software logs by the literature~\cite{xu2002prefix, li2005canine, mohammady2018preserving}. For example,the  IP address \textit{`129.188.154.200'} can reveal that it belongs to Motorola Inc. and originates from Schaumburg, Illinois, USA. While an IP address may not be traced directly to an individual, it can often be linked to a household or company, which is typically considered sensitive. Similarly, a hostname can provide insights into the user's location. For instance, the hostname \textit{`ec2-52-80-34-196.cn-north-1.compute.amazonaws.com.cn`} indicates that the host machine is utilizing Amazon services, with the \textit{`cn-north-1`} label suggesting that it is located in Beijing, China. In another example, the hostname \textit{`proxy.cse.cuhk.edu.hk`} reveals that it belongs to the \textit{Computer Science and Engineering (CSE)} department at the \textit{Chinese University of Hong Kong (CUHK)}.

One major category of potentially sensitive data that has been less emphasized in studies is the configuration details (see Section~\ref{sec:articles_tools}). We define configuration details as all information related to system settings and configurations. For instance, in the Thunderbird dataset, we encounter information such as the CPU specification \textit{(Intel(R) Xeon(TM) CPU 3.60GHz stepping 03)}, the number of CPU cores \textit{(checking TSC synchronization across 4 CPUs: passed)}, memory capacity \textit{(L2 cache: 2048K)}, and network connection status \textit{(NIC Link is Up 1000 Mbps Full Duplex)}. Similarly, the Linux dataset includes details such as CPU specification \textit{(CPU: Intel Pentium III)}, Linux version \textit{(Linux version 2.6.5-1.358)}, and gcc version \textit{(gcc version 3.3.3 20040412)}. Another example from \textit{``Jul 27 14:41:59 combo kernel: PCI: Using IRQ router PIIX/ICH [8086/2410] at 0000:00:1f.0''} reveals that the chipset is manufactured by Intel based on \textit{PIIX/ICH} and vendor ID \textit{8086}. Additionally, we find configuration details such as the number of nodes, sleep/timeout durations, CPU fan speed, and installed package names in Android device logs. We believe that this information could be sensitive to the log provider and pose a risk of unintentional data leakage, for example, exposing detailed information about a company's computing environment that may be vulnerable to adversarial attacks. Configuration details are found in 36\% of log datasets.

Most of the log datasets examined are raw and unaltered. According to the description of the LogHub collection, ``Wherever possible, the logs are NOT sanitized, anonymized, or modified in any way''. Further confirmation from the LogHub collection publishers verified that most datasets remain intact, with the exceptions of Android, HDFS, and Windows datasets, which were edited to remove certain private IP addresses and package names. For the ITA collection, the Calgary dataset has had IP addresses entirely removed for privacy reasons, file paths shortened, and file names changed. In the WorldCup98 and SDSC datasets, IP addresses have been renumbered and converted to integer values. In the Boston dataset, User IDs have been transformed using a one-way function.

\begin{mybox}
Timestamps, IP addresses, file paths, IDs, and components are the most common attributes in logs. We also identify attributes such as configuration details and environmental data.
\end{mybox}
\section{Privacy Regulations}
\label{sec:regulations}

Privacy regulations are designed to protect personal data and ensure that organizations handle this data responsibly. Different countries and continents have their own sets of privacy regulations, with some countries having multiple regulations to address various aspects of data protection. For example, the United States has \textit{Health Insurance Portability and Accountability Act (HIPAA)} for health data privacy and \textit{California Consumer Privacy Act (CCPA)} for consumer data protection. In this study, we focus on some of the most widely recognized and adopted regulations, including the \textit{General Data Protection Regulation (GDPR)} in Europe, HIPAA, and CCPA in the United States, \textit{Personal Information Protection and Electronic Documents Act (PIPEDA)} in Canada, and ISO standard 27001, which provides an international framework for information security management. By analyzing these key regulations, we aim to cover a broad range of  privacy practices and requirements.

\subsection{GDPR}

GDPR~\cite{gdpr} is a data regulation on information privacy established by the \textit{European Union (EU)} that applies to the \textit{European Economic Area (EEA)}. Established in 2018, GDPR is widely recognized as one of the most comprehensive and influential data privacy regulations globally. It sets strict guidelines for the collection, processing, storage, and transfer of personal data, aiming to protect the privacy and rights of individuals within the EU and EEA.

At the core of GDPR is the broad definition of personal data, which encompasses \textbf{any information related to an identified or identifiable natural person}. This definition is wide-ranging, covering both direct identifiers, such as names and identification numbers, and indirect identifiers, such as location data and IP addresses. The regulation emphasizes that IP addresses can be considered personal data if they can be linked to an individual, either directly or indirectly. For instance, if a data controller has the means to obtain additional information that could identify the person behind an IP address, that IP address would be classified as personal data under GDPR. Moreover, GDPR acknowledges that identifiers from devices, applications, and protocols can build detailed profiles of individuals. These identifiers, such as IP addresses and cookie IDs, can leave digital footprints that, when combined with other data, might disclose an individual's identity.

In the famous case of ``Breyer v. Germany''~\cite{breyer}, the \textit{Court of Justice of the European Union (CJEU)} examined whether a dynamic IP address qualifies as personal data. The court ruled that while a dynamic IP address on its own might not identify an individual, it can be considered personal data if it can be combined with additional information that allows for identification. This is particularly relevant when an online service provider has the legal ability to access such identifying information. 

\subsection{HIPAA}

HIPAA~\cite{hipaa} is a U.S. legislation aimed at protecting the privacy and security of individuals' health information. Established in 1996, HIPAA creates national standards for data protection, particularly focusing on the privacy of health information. This regulation is crucial in the healthcare industry, ensuring that sensitive patient information is adequately protected while allowing for the necessary flow of data to provide high-quality healthcare services. Unlike GDPR, which broadly defines personal data, HIPAA specifies 18 sensitive identifiers that must be protected. These identifiers range from direct identifiers such as names, phone numbers, and Social Security numbers, to indirect identifiers such as geographic subdivisions, dates related to an individual, and IP addresses. Some of these 18 identifiers could be found in software logs, \textbf{such as dates, IP addresses, device identifiers, URLs, and any other unique identifying number, characteristic, or code.}

\subsection{CCPA}

CCPA~\cite{ccpa} is a state statute intended to strengthen privacy rights and consumer protection for residents of California, United States. Signed into law in 2018, CCPA provides California residents with significant control over their personal information, granting them the right to know what data is being collected about them, the right to request the deletion of their data, and the right to opt out of the sale of their personal information. Under the CCPA, personal information is broadly defined as \textbf{any data that identifies, relates to, or could reasonably be linked with an individual or their household}. This includes obvious identifiers such as names, social security numbers, and email addresses, but it also extends to less direct identifiers such as internet browsing history and geolocation data. Importantly, CCPA distinguishes between ``personal information'' and ``sensitive personal information'', which includes highly specific data such as government identifiers, precise geolocation, email contents, genetic data, and biometric identifiers. The CCPA recognizes that not all personal information requires the same level of protection, including exemptions for publicly available government records and data already governed by other laws such as HIPAA. 

\subsection{PIPEDA}

PIPEDA~\cite{pipeda} is a Canadian law that regulates how private sector organizations handle personal information during commercial activities. Established in 2000, PIPEDA mandates that organizations collect, use, and disclose personal information responsibly, ensuring the privacy of individuals. Personal information under PIPEDA includes \textbf{any factual or subjective information, recorded or not, about an identifiable individual}, such as age, name, income, ID numbers, or opinions. This broad definition encompasses various forms of data, making PIPEDA a comprehensive privacy law. However, PIPEDA does not apply universally. Exceptions include personal information managed by federal government organizations under the Privacy Act, provincial and territorial governments, and business contact information used solely for professional communications. Additionally, PIPEDA does not cover information collected for personal, journalistic, artistic, or literary purposes. 

\subsection{ISO27001}

ISO 27001~\cite{iso27001} is an internationally recognized standard for \textit{Information Security Management Systems (ISMS)}, providing \textbf{a framework for organizations to systematically manage sensitive data.} Unlike previously discussed regulations, ISO 27001 does not define personal or sensitive data but focuses on managing information security risks, allowing organizations to determine sensitivity based on context and requirements. Initially published in 2005, ISO 27001 guides the implementation and maintenance of an ISMS through a combination of technological, organizational, physical, and human controls, offering flexibility for different industries and data types.

\begin{mybox}
While numerous data privacy regulations exist, none specifically address software logs. Therefore, it is essential to extract relevant information from these regulations that could be applicable to logs. Some regulations explicitly define personal data and specify attributes that need protection. For instance, GDPR and HIPAA both classify IP addresses as sensitive data. In contrast, ISO 27001 does not define personal or sensitive data but instead offers a flexible framework for managing any data that an organization identifies as sensitive.
\end{mybox}

\section{Articles and Tools}
\label{sec:articles_tools}

In this section, we conduct a \textit{Systematic Literature Review (SLR)} of existing research articles and tools related to log privacy to deepen our understanding of privacy in software logs. The main goal of this review is to determine which software log attributes are regarded as sensitive by researchers.

\subsection{Approach} 
We conduct an extensive search across two research databases, initially identifying 88 potential candidate articles. Applying inclusion and exclusion criteria, we reduce the selection to 22 papers. To ensure a thorough review, we perform forward and backward snowballing, leading to the inclusion of an additional 36 articles. Combining these, we analyze a total of 58 papers. Our selection process follows the systematic approach recommended by Wohlin et al.~\cite{wohlin2012experimentation}, as detailed below.

\subsubsection{Searching Research Databases} 
\label{sec:search_databases}
We conduct our literature search using two well-known databases: the IEEE Xplore Digital Library\footnote{\url{https://ieeexplore.ieee.org/}} and the ACM Digital Library.\footnote{\url{https://dl.acm.org/}} Our objective is to find papers from both conference proceedings and journals. To identify articles that focus on defining privacy in software logs or perform log anonymization, we employ the following search query:

\textit{Title: (log AND privacy) OR (log AND $sensitiv^*$) OR (log AND $anonymiz^*$)}

Using this query, we find 57 relevant articles from the IEEE library. We also obtain an additional 31 articles from the ACM library after deduplication. In total, combining results from both databases, we identify 88 candidate articles for further review.

\subsubsection{Article Selection} 
\label{sec:article_selection}
We start by reviewing each paper's abstract and continue through the entire article, applying the inclusion and exclusion criteria until a final decision is reached.

\vspace{0.4em} \textit{Inclusion criteria:}
\begin{itemize}
    \item Papers must be written in English.
    \item Papers must be in conference proceedings or journals.
    \item Papers must be in the scope of software logs.
    \item Papers must either define privacy in software logs, introduce an anonymization tool, explore privacy considerations, or improve privacy measures in software logs.
\end{itemize}

\textit{Exclusion criteria:}
\begin{itemize}
    \item Papers not publicly available.
    \item Papers that focus on general data privacy without specific emphasis on software logs.
    \item Papers working with other types of data such as search logs.
\end{itemize}

After performing the article selection step, we end up selecting 22 out of the 88 candidate articles.

\subsubsection{Forward and Backward Snowballing} 
\label{sec:snowball}

To ensure thoroughness in our study, we conduct both forward and backward snowballing on the initially selected 22 papers. In this phase, we examine the papers that have cited these selected articles (i.e., forward snowballing) as well as the references cited within them (i.e., backward snowballing). For each of these papers, we apply our article selection criteria (detailed in Section~\ref{sec:article_selection}) to determine if they meet our inclusion and exclusion criteria. This process results in the identification of 36 new articles. By combining these with the articles initially found through our database search, we compile a total of 58 papers that specifically address privacy in software logs.

\subsubsection{Information Extraction} 
\label{sec:extraction}
In the final step, we analyze how each paper defines privacy in the context of log data. We review their approach sections to determine which attributes are considered sensitive.

\subsection{Results} 
\noindent\textbf{IP addresses, timestamps, and port numbers are most frequently considered as sensitive attributes in logs.}
Table~\ref{tab:attributes_papers} summarizes the attributes most frequently identified as sensitive across the reviewed articles. Among these, IP address stands out as the most discussed attribute, appearing in 59\% of the analyzed articles. This is followed by the timestamp, which is considered sensitive in 28\% of the studies, and the port number, which is mentioned in 21\% of the papers. 

Comparing Table~\ref{tab:attributes_preliminary} with Table~\ref{tab:attributes_papers}, we find a notable overlap between the attributes found to frequently occur in software logs and those highlighted in the existing literature. Our log analysis indicates that every software log includes the timestamp attribute, with 80\% also having the IP address. This aligns with the literature, where IP address and timestamp are identified as the most frequently studied attributes. We also find that many attributes, such as port number, various IDs, usernames, and file paths, are common in both tables.

We identify three new attributes from the reviewed articles that were not present in the analyzed logs: network-related features, email addresses, and location data. Network-related attributes are features that are specific to network logs, such as the ``don't fragment'' bit, \textit{Transmission Control Protocol (TCP)} window size, and \textit{Time To Live (TTL)}. The ``don't fragment'' bit is a flag in the IP header that determines if a packet should be fragmented. TCP window size specifies the amount of data the receiver can handle at once. TTL is a field in the IP header that limits the number of hops (i.e., routers or devices) a packet can make before being discarded. The other newly detected attributes are email addresses and location information such as address or zip code. 

On the other hand, several attributes that are commonly found in software logs have not been thoroughly discussed in the literature. Some of these attributes are not mentioned in any of the reviewed articles. For instance, attributes such as component, log level, request status code, and environmental data fall into this category. These attributes can provide insights into the internal workings of a system, such as specific components involved in processing or the severity of events. Additionally, some attributes are mentioned in articles but at a relatively low frequency. For example, although file paths are present in 72\% of the logs examined, they are discussed in only 7\% of the articles. Similarly, while 44\% of logs contain hostnames, only 5\% of studied articles discuss this.

Attributes that have been discussed in only one article are categorized under ``others.'' This group includes computer commands, passwords, URLs, ages, and credit cards.

\begin{table}[t]
\centering
\caption{Usage of sensitive log attributes in reviewed articles}
\footnotesize
\resizebox{0.8\textwidth}{!}{%
\begin{tabular} {|p{0.18\textwidth} |p{0.04\textwidth} |p{0.18\textwidth} |p{0.04\textwidth} |p{0.18\textwidth} |p{0.04\textwidth}|}
\hline
\textbf{Attribute} & \textbf{Freq. (\%)} & \textbf{Attribute} & \textbf{Freq. (\%)} & \textbf{Attribute} & \textbf{Freq. (\%)}  \\ \hline
IP address & 59 & Username & 14 & Email & 7 \\ \hline
Timestamp & 28 & Request response size & 10 & File path & 7 \\ \hline
Port number & 21 & Configuration details & 9 & Hostname & 5 \\ \hline
IDs & 17 & MAC address & 9 & Location & 3 \\ \hline
Network-related & 16 & Request protocol & 9 & Others & 9 \\ \hline
\end{tabular}%
}
\label{tab:attributes_papers}
\end{table}

\noindent\textbf{Many studies only focus on the privacy of IP addresses.}
14 of the reviewed articles (almost 24\%) focus exclusively on IP addresses as the sensitive attribute (e.g.,~\cite{xu2002prefix, han2020aft, manocchio2024configurable, mohammady2018preserving}). One of the earliest and most influential works in software log privacy by Xu et al.~\cite{xu2002prefix}, introduces CryptoPAn, a tool for prefix-preserving IP address anonymization. This means that if two original IP addresses share a k-bit prefix, their anonymized versions will also share the same k-bit prefix. CryptoPAn employs a cryptography-based approach, using bitwise anonymization, where the anonymized IP addresses are determined by previous anonymization. However, CryptoPAn is known to be vulnerable to fingerprinting and injection attacks~\cite{brekne2005anonymization, brekne2005circumventing}, where attackers use known network flows or inject fake flows to infer other flows by analyzing unchanged fields such as timestamps.

In a more recent development, Han et al.~\cite{han2020aft} introduce AFT-Anon, a method designed for real-time anonymization of IP addresses in network packets. This approach leverages flow tables to ensure efficient and secure anonymization as data flows through the network. In another study, Manocchio et al.~\cite{manocchio2024configurable} explore the use of a non-reversible pseudonymization function to anonymize IP addresses. They compare the accuracy of two machine learning models on a network dataset both before and after anonymization. Their findings reveal a decrease in accuracy, ranging from 1\% to 4\%, post-anonymization, highlighting the trade-off between privacy and data utility.

In a different approach, Mohammady et al.~\cite{mohammady2018preserving} propose the Multi-view method, generating multiple anonymized versions of network traces that are so similar adversaries cannot distinguish between them, preserving privacy. Simultaneously, one version provides accurate analysis results for the data owner, ensuring utility. However, the study recommends creating 160 distinct data views for the analyst, which may not be practical due to time and computational constraints.

\noindent\textbf{Many studies focus only on the network-related attributes.} As discussed in Section~\ref{sec:preliminary_study}, software logs can originate from various sources, including distributed systems, supercomputers, operating systems, and server applications. Despite this broad spectrum of software logs, many selected articles, and particularly most anonymization tools, focus exclusively on network logs (e.g.,~\cite{li2005canine, slagell2006flaim}).

As highlighted by Slagell et al.~\cite{slagell2005sharing}, there are 16 types of network logs that require anonymization, including TCPdump, NetFlow, and Syslog. Several well-known log anonymization tools are designed to handle these network data types. For instance, CANINE, developed by Li et al.~\cite{li2005canine}, is a tool specifically for anonymizing NetFlow data. CANINE can anonymize five different attributes, including IP addresses, timestamps, port numbers, request protocols, and response sizes. It employs different anonymization techniques for each attribute; for instance, IP addresses can be anonymized using truncation, random permutation, or prefix-preserving pseudonymization.

FLAIM is widely regarded as one of the most comprehensive log anonymization tools available. Introduced by Slagell et al.~\cite{slagell2006flaim}, FLAIM employs a range of anonymization techniques, including black marker, truncation, and hashing, to anonymize netfilter logs, pcap traces, and NetFlow data, covering attributes such as IP addresses, timestamps, MAC addresses, host names, and port numbers. Additionally, it can handle specific network-related attributes such as \textit{Internet Control Message Protocol (ICMP)} codes and types, the ``don't fragment'' bit, TCP window size, and TCP options.

AnonTool is a specialized anonymization tool designed for NetFlow data~\cite{foukarakis2009deep, foukarakis2007flexible}. Developed in C, it can anonymize live packet traces in the libpcap file format, which is used for network traffic capture. AnonTool handles various attributes, including IP addresses, timestamps, port numbers, response sizes, and specific NetFlow attributes such as \textit{Type of Service (TOS)} and TCP flags.

Created by Minshall, TCPdpriv~\cite{tcpdpriv} is another anonymization tool that works with the libpcap library. However, it has some limitations, including compatibility with only SunOS, Solaris, and FreeBSD systems. Based on TCPdpriv, Plonka developed IP2anonIP~\cite{ip2anonip} to convert IP addresses into hostnames or anonymous IP addresses. IP2anonIP also offers the option to add custom fields, but preparing a dataset for a single day can take up to several hours.

\noindent\textbf{Some studies discuss privacy without specifying any attribute.} 13 articles (22\%) do not identify any log attributes as sensitive. We categorize these articles into two main groups: (1) surveys and taxonomies, and (2) log parsers.

\textit{1) Surveys and taxonomies.}
Burkhart et al.~\cite{burkhart2010role} explore the vulnerabilities of network trace anonymization techniques, particularly focusing on the threat posed by traffic injection attacks. The authors demonstrate the ease of executing these attacks and discuss why some anonymization techniques such as aggregation, randomization, or field
generalization might be ineffective. They argue that anonymization alone is insufficient for data protection and must be integrated with legal, social, and technical measures to ensure effective network trace sharing and data protection. 

King et al.~\cite{king2009taxonomy} present a taxonomy and adversarial model for classifying attacks against network log anonymization. By categorizing various attacks, they provide a framework for identifying the strengths and weaknesses of different anonymization techniques, including pseudonymization, IP address prefix preserving, black marker, and random permutation.

Silva et al.~\cite{silva2021privacy} provide a survey focused on privacy within cloud computing. The paper reviews the current state of privacy in cloud services, addressing various threats, concepts, and technologies. The authors examine \textit{Privacy Enhancing Technologies (PETs)} and different anonymization mechanisms, tools, models, and metrics, exploring their relevance to cloud environments. The paper highlights the importance of integrating various privacy mechanisms and metrics to ensure compliance with regional regulations and strengthen data protection in the cloud.

In another literature survey, Majeed et al.~\cite{majeed2022toward} emphasize the increasing importance of privacy protection in data sharing, particularly through \textit{Clustering-based Anonymization Mechanisms (CAMs)}. CAMs are identified as effective methods for preserving both privacy and utility in data publishing, surpassing traditional models such as k-anonymity and differential privacy. The paper analyzes existing CAMs, categorizing them by data types and assessing their strengths and weaknesses. It also discusses the challenges CAMs encounter, such as managing heterogeneous data, ensuring resilience against AI-powered attacks, and balancing privacy with utility.

\textit{2) Log parsers.} A trend appears to be emerging in designing log parsers that prioritize privacy concerns. The Delog parser~\cite{agrawal2019delog} is designed to enhance privacy and performance in log filtering for long-running software applications. It leverages previous successful runs to automatically identify errors, applies \textit{Locality Sensitive Hashing (LSH)} to parse log lines, and employs a privacy-preserving mechanism, using encryption and bloom filters, to securely learn new patterns from client logs. The parser is validated using four of the LogHub datasets, namely HDFS, BGL, HPC, and Zookeeper.

In another attempt, Li et al.~\cite{li2023triplelp} introduce TripleLP, a blockchain-based tool for privacy-focused log management. TripleLP automates the log parsing process by organizing logs into a structured format within a blockchain, facilitating efficient analysis and secure storage. It consists of three main components: the log server, log parser, and blockchain. The log parser dynamically adjusts templates and parameters, using a heuristic log tree approach to classify and analyze logs. After parsing, logs are encrypted and verified via blockchain before being stored securely. The tool is validated using four of the LogHub datasets: HDFS, Zookeeper, Thunderbird, and HealthApp.

\begin{mybox}
A systematic review of 58 articles reveals that IP addresses, timestamps, and port numbers are the most frequently studied sensitive attributes in logs. Some research also highlights additional sensitive elements, such as network-related attributes (e.g., TCP window size), email addresses, and location data, though they may not appear in many types of software logs. We also find that most studies only focus on a small set of attributes, with only a few addressing multiple sensitive attributes.
\end{mybox}

\section{Industry Practices}
\label{sec:industry}

In this section, we design a survey to understand industry professionals' perspectives on software log privacy. The survey explores which attributes in software logs are considered sensitive by those involved in log management and anonymization. By gathering insights from experts, we aim to complement our findings from log analysis, regulatory review, and literature research, providing an overview of prioritized attributes and handling approaches for privacy protection.

\subsection{Survey Design} In designing the survey, we use a variety of question types, including yes/no questions, multiple-choice questions, Likert-scale questions, and open-ended questions, to capture both quantitative and qualitative insights from participants. Using these different types of questions, we aim to gather diverse data that includes both statistical and descriptive insights, allowing us to analyze not just what professionals think but also how they justify their opinions and approaches.

The survey is designed to respect the privacy and time of the participants. It takes between 5-10 minutes to complete, ensuring that it is not inconvenient. We do not collect any personal information, such as names, email addresses, or the company names where participants are employed, ensuring that responses remain anonymous and protecting participants' confidentiality.

Conducting a pilot survey is beneficial in order to identify and address any unforeseen problems and challenges before releasing the final version. It is crucial for pilot studies to use the same artifacts and procedures (e.g., the invitation, explanation, and format) as those used in the main study~\cite{kitchenham2015evidence}. Therefore, we conducted a pilot survey with the first three respondents. Based on their feedback, we made slight changes in the wording of some questions and answer options to improve clarity and ensure that the questions are easily understood by the target audience.

\subsection{Participant Selection Strategy} According to Kitchenham et al.~\cite{kitchenham2008personal}, it is crucial that the target population is capable of addressing the research questions effectively. To identify the relevant audience for software log privacy, we concentrate on professionals involved in log management, log analysis, and data privacy. We select survey participants using two methods. The first method involves searching for specific roles and keywords on LinkedIn,\footnote{\url{https://www.linkedin.com/}} a popular professional networking platform. We use four keywords: (1) privacy engineer, (2) data privacy officer, (3) IT manager, and (4) IT security engineer. These keywords help us filter professionals likely to have direct experience with software logs and data privacy. The second method involves distributing the survey link through the authors' network to reach industry partners and professionals. Due to the broad distribution, we lack the exact count of recipients, preventing us from calculating a precise participation rate.

We conduct the survey over three weeks, providing sufficient time for participants to respond while ensuring timely data collection. This duration is chosen to strike a balance between obtaining a robust sample size and accommodating the busy schedules of professionals. 

\begin{figure}
\centering
\includegraphics[width=1\textwidth]{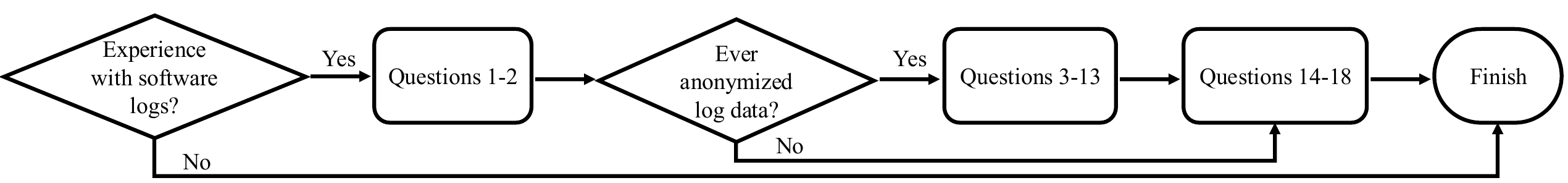}
\caption{Structure of our survey}
\label{fig:survey_structure}    
\end{figure}

\subsection{Survey Questions}
Figure~\ref{fig:survey_structure} shows the survey flowchart, which starts by determining if the participant has experience in analyzing, processing, or managing software logs. If the answer is ``no'', the participant is ineligible to answer the survey and does not proceed further; if the answer is ``yes'', the participant moves on to answer questions 1 and 2.

The survey then asks if participants have experience with log anonymization. If the response is ``no'', the participant skips directly to question 14. On the other hand, if the participant has experience in log anonymization, they proceed to question 3. This branching structure effectively creates two participant profiles: one for those with software log experience but no anonymization, and another for those experienced in both. Questions 3 to 13 focus more on specific aspects of log anonymization, such as balancing utility with privacy, to better understand the practices and challenges faced by professionals in the field. Below, we provide our survey questions.

\textit{Multiple-choice questions:} \\
Q1. Are you familiar with the following data protection regulations and standards? \\
Q2. Which of the following attributes do you consider sensitive in software logs? \\
Q3. When you want to share software logs, which attributes do you anonymize? \\
Q4. Why does your response to Q2 differ from Q3? \\
Q5. What factors influence your decision to anonymize certain log attributes? \\
Q6. Which anonymization techniques do you primarily use for log data? 

\textit{Likert-scale questions:} \\
Q7. How much do you believe anonymization impacts the usefulness of the data? \\
Q8. How challenging do you find balancing anonymization with preserving the utility of log data? \\
Q9. How effective are your anonymization practices in protecting sensitive information? \\
Q10. How effective are your anonymization practices in preserving utility? \\
Q11. How efficient are your anonymization practices regarding computation cost and time? 

\textit{Open-ended questions:} \\
Q12. Can you share strategies or best practices to maintain data utility while anonymizing logs? \\
Q13. What challenges do you face when applying anonymization techniques to software logs? \\
Q14. Do you have any additional comments, suggestions, or insights?

\textit{Demographic questions:} \\
Q15. Which best describes your role in the organization? \\
Q16. How many years of experience do you have? \\
Q17. In which industry does your organization operate? \\
Q18. What is the approximate size of your organization?

We incorporate insights from previous sections when designing the survey. For example, we use the regulations discussed in Section~\ref{sec:regulations} to form the response options for Q1. Similarly, we adapt Table~\ref{tab:attributes_preliminary} (with some minor modifications) as response options for Q2 and Q3. 

\subsection{Participants Demographics}
During the three weeks the survey was open, we collected 61 responses. Among them, 16 respondents lacked experience with software logs and were therefore ineligible to complete the survey, leaving 45 valid participants. Of these, 53\% fell into the first profile, having experience with software logs but not with log anonymization, while the remaining 47\% belonged to the second category, having experience with both software logs and log anonymization. Out of our 45 participants, only seven were unfamiliar with any data regulations. The remaining 84\% were familiar with at least one, with all aware of the GDPR, making it the most recognized data regulation. Other regulations, such as CCPA (58\%), HIPAA (53\%), and PIPEDA (49\%), were also well-known. Additionally, 29\% of participants mentioned familiarity with regulations not covered in this study, such as Brazil's \textit{General Personal Data Protection Law (LGPD)} or various provincial regulations. Based on Q15 responses, we categorize participants' roles into six groups to better understand their professional backgrounds and expertise. These categories are \textbf{Data Privacy roles}, \textbf{Software Engineering roles}, \textbf{Security roles}, \textbf{Network/System roles}, \textbf{Data Science/Engineering roles}, and \textbf{Management roles}. 


\begin{table*}[t]
    \caption{Demographics of survey participants}
    \centering
    \footnotesize

    \begin{subtable}[t]{0.45\textwidth}
        \centering
        \caption{Job Role}
        \begin{tabular}{lc}
            \toprule
            Job Role & Percentage \\
            \midrule
            Data Privacy roles & 40.0\% \\
            Software Engineering roles & 24.4\% \\
            Security roles & 20.0\% \\
            Network/System roles & 6.8\% \\
            Data Science/Engineering roles & 4.4\% \\
            Management roles & 4.4\% \\
            \bottomrule
        \end{tabular}
    \end{subtable}
    \hspace{0.05\textwidth}
    \begin{subtable}[t]{0.45\textwidth}
        \centering
        \caption{Experience}
        \begin{tabular}{lc}
            \toprule
            Experience & Percentage \\
            \midrule
            Less than 1 year & 4.5\% \\
            1-3 years        & 11.1\% \\
            4-6 years        & 20.0\% \\
            7-10 years       & 22.2\% \\
            More than 10 years & 42.2\% \\
            \bottomrule
        \end{tabular}
    \end{subtable}
    
    \vspace{0.5cm} 

    \begin{subtable}[t]{0.45\textwidth}
        \centering
        \caption{Industry}
        \begin{tabular}{lc}
            \toprule
            Industry & Percentage \\
            \midrule
            Technology & 69.0\% \\
            Finance & 8.9\% \\
            Healthcare & 4.4\% \\
            Manufacturing & 4.4\% \\
            Government & 4.4\% \\
            Other & 8.9\% \\
            \bottomrule
        \end{tabular}
    \end{subtable}
    \hspace{0.05\textwidth}
    \begin{subtable}[t]{0.45\textwidth}
        \centering
        \caption{Organization Size}
        \begin{tabular}{lc}
            \toprule
            Size & Percentage \\
            \midrule
            1-100 employees  & 17.8\% \\
            101-500 employees & 4.4\% \\
            More than 500 employees  & 77.8\% \\
            \bottomrule
        \end{tabular}
    \end{subtable}
    \label{tab:demographics}
\end{table*}

\textbf{Data Privacy roles} include professionals focused on ensuring compliance with data protection regulations and implementing privacy-preserving practices, such as data privacy officers and privacy engineers. \textbf{Software Engineering roles} consist of developers involved in designing, building, and maintaining software systems and applications. \textbf{Security roles} are experts responsible for protecting systems and data from unauthorized access, breaches, and other cyber threats, such as security engineers. \textbf{Network/System roles} include specialists in designing, managing, and optimizing IT infrastructure, such as network architects. \textbf{Data Science/Engineering roles} involve professionals who analyze data, build models, and develop data pipelines to derive insights. Lastly, \textbf{Management roles} include senior leaders and team leads who oversee technical teams, make strategic decisions, and manage organizational resources, such as a \textit{Chief Technology Officer (CTO)}.

Table \ref{tab:demographics} shows the demographics of survey participants, highlighting their job roles, experience, industries, and organization sizes. Most respondents hold roles related to data privacy (40.0\%), software engineering (24.4\%), and security (20.0\%). Most participants are highly experienced in their fields; 42.2\% have over 10 years of experience, and 22.2\% have between 7 to 10 years, indicating senior-level expertise. The majority work in the technology industry (69.0\%), closely tied to software logs and data privacy. Additionally, 77.8\% of the participants are employed in large organizations with more than 500 employees, which often have more complex data privacy and security challenges compared to smaller companies. Overall, the combination of extensive experience, professional focus on relevant fields, and employment in large organizations suggests that our survey results can provide valuable insights into industry practices and perspectives on log privacy.

\subsection{Survey Results}
\noindent\textbf{IP addresses, MAC addresses, host names, and file paths are the most sensitive log attributes from the industry perspective.}
A significant majority of industry professionals, 87\%, regard IP addresses as sensitive data in software logs that require anonymization, followed closely by MAC addresses at 82\%. These two attributes are considered the most critical to protect, far higher than other attributes. Over 50\% of the participants also identify hostnames and file paths as sensitive elements requiring careful handling. Other attributes considered sensitive by industry include various forms of IDs, URLs, port numbers, components, and usernames. 

9\% of participants propose additional factors in determining sensitivity in software logs, emphasizing the importance of contextual and legal considerations. For instance, Respondent 30 notes that ``Sensitivity depends on the context of the environment, community, and secondary actions (e.g., re-identification, linking, and correlation.)'' Similarly, Respondent 31 mentions that ``IP may be sensitive or public, other elements such as IDs, date and time, and user actions associated with a log may be sensitive depending on the context in which it is used, how it is connected to a person or represents a person''. Respondent 46 also points out that ``Sensitivity could depend on the type of website visited (i.e., healthcare vs. gaming)''. Additionally, some respondents highlight that sensitivity should be assessed based on relevant laws and regulations.

\begin{table}[t]
\centering
\caption{Sensitive log attributes from industry perspective}
\footnotesize
\resizebox{0.8\textwidth}{!}{%
\begin{tabular} {|p{0.18\textwidth} |p{0.04\textwidth} |p{0.18\textwidth} |p{0.04\textwidth} |p{0.18\textwidth} |p{0.04\textwidth}|}
\hline
\textbf{Attribute} & \textbf{Freq. (\%)} & \textbf{Attribute} & \textbf{Freq. (\%)} & \textbf{Attribute} & \textbf{Freq. (\%)}  \\ \hline
IP address & 86 & Component & 27 & Request method & 9 \\ \hline
MAC address & 82 & Username & 20 & Request status code & 9 \\ \hline
Hostname & 59 & Configuration details & 18 & Request response time & 4 \\ \hline
File path & 52 & Date and Time & 18 & Request response size & 2 \\ \hline
IDs & 43 & Environmental data & 11 & None & 2 \\ \hline
URL & 39 & LOG level & 11 & Others & 9 \\ \hline
Port number & 34 & Request protocol & 9 \\ \cline{1-4}
\end{tabular}%
}
\label{tab:attributes_industry}
\end{table}

\noindent\textbf{Log utility, policies, and technical challenges are among the main factors influencing the decision for anonymizing a log attribute.} The process of deciding whether to anonymize a log attribute varies widely among professionals in the industry. Through Q4, we explore why there is an inconsistency between what respondents consider sensitive and what they actually anonymize in practice. Many participants cite various practical reasons: 52.4\% point to the potential utility loss, 38.1\% anonymize only as required by policies or regulations, 28.6\% face technical limitations, and a smaller group, 14.3\%, anonymize fewer attributes because their logs are not shared externally.

When asked what factors influence their decision to anonymize specific log attributes (Q5), most participants highlight legal compliance (81\%) and the risk of re-identification (81\%) as key factors. Company policies also play a significant role, with 76.2\% of respondents noting it. Additionally, 55\% of participants mention the impact on data utility as an influencing factor. Customer requirements are less frequently cited, with 42.9\% indicating this as a consideration. These responses highlight a complex decision-making environment where legal obligations, technical feasibility, data utility, and organizational policies all play a role in shaping anonymization strategies for software logs.

\noindent\textbf{Finding a balance between privacy and utility is extremely challenging.} Finding a balance between privacy and utility in log anonymization is a significant challenge, as highlighted by the survey responses. A notable number of respondents recognize the impact of anonymization on data utility, with 33.3\% reporting a ``moderate'' impact and another 38.1\% indicating a ``significant'' impact. This indicates that while anonymization is essential for protecting sensitive information, it often comes at a cost to data usability. This trade-off emphasizes the difficulty practitioners face in balancing to protect privacy while maintaining data utility for analytical purposes.

Respondents also note the complexity of balancing privacy and utility, with 76.1\% rating it as ``extremely challenging'', ``challenging'', or ``moderately challenging''. This reveals the need for more effective anonymization methods that can better preserve both privacy and utility. Current techniques such as differential privacy, pseudonymization, and data masking offer varied levels of success. For instance, while differential privacy and synthetic data generation can enhance privacy without completely sacrificing data utility, pseudonymization allows for conditional re-identification when necessary. These techniques reflect the need for context-specific approaches. 

Despite using various techniques, only 57.1\% of respondents consider their anonymization practices ``effective'' or ``very effective'' in protecting sensitive information. Similarly, only 57.2\% believe their methods adequately maintain data utility. This indicates that, although there is some confidence in current anonymization strategies, there is still considerable room for improvement. The diversity in anonymization approaches highlights the need for more sophisticated and adaptive solutions capable of handling nuances of different data types and use cases. Developing frameworks that dynamically adjust to changing requirements and contexts could significantly enhance the effectiveness of anonymization practices in balancing privacy and utility.

\noindent\textbf{Current anonymization practices face efficiency, technical, and operational challenges.} A significant finding from the survey is that only 33.3\% of the participants view their anonymization techniques as efficient in terms of computational costs and time, with the remaining 66.7\% categorizing them as either ``moderately efficient'' or ``not efficient''. None of the respondents rated their methods as ``highly efficient'', highlighting a clear need for optimization in this area. 

Participants also highlighted several technical challenges in anonymization, such as maintaining consistency and referential integrity, especially with high-volume, real-time logs that are often unstructured. The variety in log formats and the need for diverse data handling tools complicate the anonymization process. Besides, many respondents pointed out that manual processes and the necessity for ongoing checks to ensure anonymization effectiveness increase both the time and computational cost while also introducing the risk of human error. This emphasizes the need for more advanced, automated solutions to manage diverse data types and volumes efficiently.

Another significant operational challenge is navigating legal and compliance complexities. Respondents frequently mentioned difficulties in aligning log anonymization practices with various regulations such as GDPR and CCPA. Balancing regulatory compliance with the need to maintain data utility for purposes such as fraud detection and security monitoring adds complexity to the process. This highlights the need for a comprehensive log anonymization approach that addresses both technical and legal dimensions, ensuring anonymization practices align with regulations.
 
\noindent\textbf{Context is important.} The participants' responses also highlight that effective anonymization practices are highly context-sensitive. For instance, respondent 47 mentions: ``Context is key! Who are you sharing the data with? What outside context do they have access to that could lead to identification? What is the threat/worst case scenario?'' Another participant mentions ``Defining PII requires interpretation of context. A log that demonstrates action associated with a person, while factual, may be PII if it demonstrates a failure to comply with legislation or policy.'' Besides, for some scenarios and for specific organizations, there might be the need for re-identification: ``There is a need, from time to time, to be able to be specific for outliers (e.g. for criminal investigations or cyber incidents). It's important to be able to have a structured approach to reidentify materials.'' This need for context-specific and adaptable anonymization approaches indicates that there is no one-size-fits-all solution for anonymization; instead, log privacy management should be dynamic, with strategies regularly evaluated and modified to address emerging requirements.

\begin{mybox}
Surveying 45 industry experts, we discovered that IP addresses, MAC addresses, host names, and file paths are most often viewed as sensitive in software logs from the industry perspective. However, multiple factors, such as log utility, policies, and technical challenges, influence practitioners' decisions to anonymize specific attributes. In particular, balancing data privacy with utility poses a significant challenge. Our findings also suggest that current anonymization tools need to improve their efficiency concerning computational costs and processing time, especially given the diverse and unstructured nature of log data.
\end{mybox}
\section{Discussion}
\label{sec:discussion}

\subsection{What Should be Anonymized in Software Logs?}

Based on the results in Section~\ref{sec:industry}, the context in which software logs are generated and shared is crucial in determining what needs to be anonymized. It is essential to understand both internal organizational policies and regional regulations. Furthermore, knowing whether the data will be shared internally among employees, externally with third parties, or even publicly is vital. All these factors must be carefully considered when deciding which log attributes require anonymization.

\begin{figure}
\centering
\includegraphics[width=0.7\textwidth]{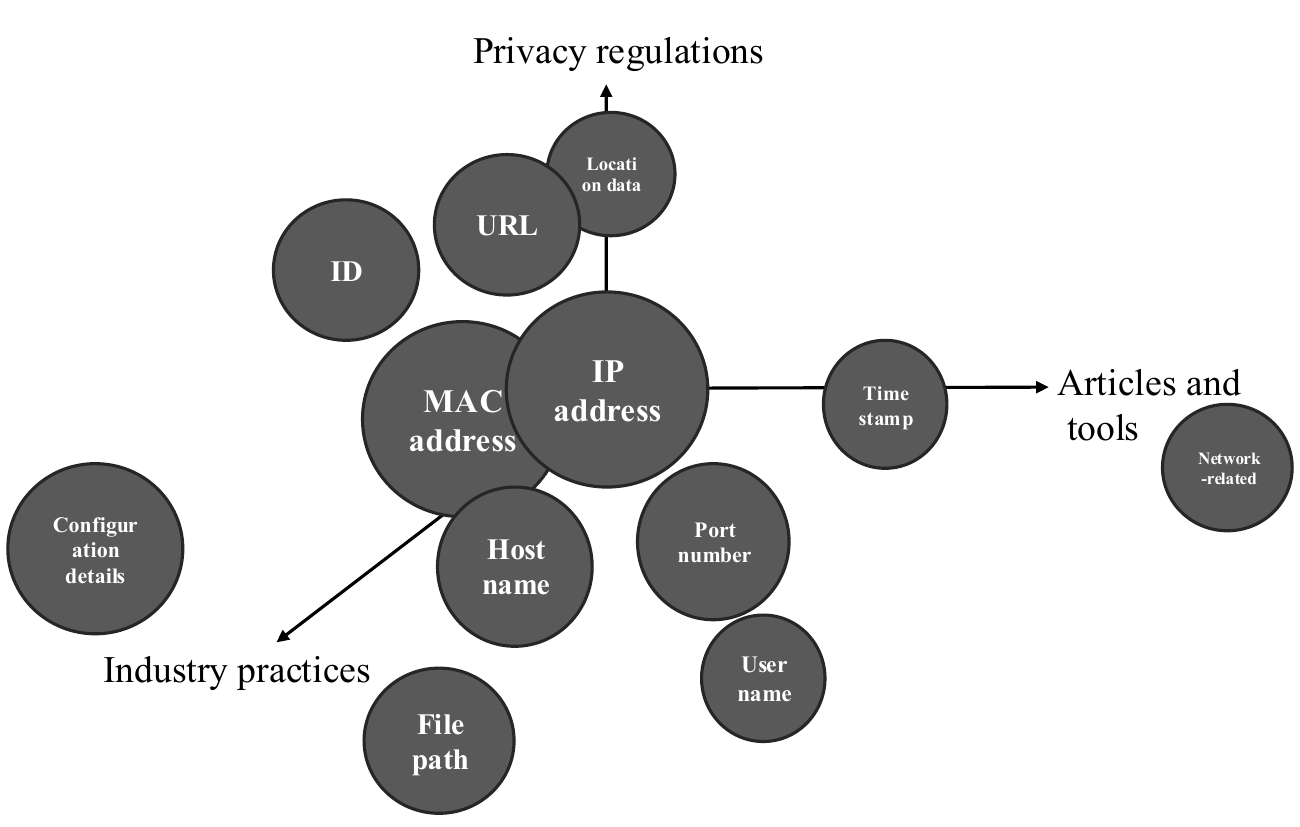}
\caption{Sensitive log attributes from different perspectives}
\label{fig:bubble}    
\end{figure}

Figure~\ref{fig:bubble} presents the sensitive log attributes from three different perspectives: privacy regulations, articles and tools, and industry practices. From an industry perspective, IP addresses, MAC addresses, hostnames, file paths, and various IDs are recognized as the most sensitive attributes in software logs. These attributes are frequently cited by experts and commonly appear in publicly available logs. For example, 86\% of survey respondents view IP addresses as sensitive, and they are found in 80\% of software logs. While IP addresses have been well-researched, other attributes are less studied. For instance, although 52\% of industry participants consider file paths sensitive, and they are present in 72\% of analyzed logs, they are rarely discussed in the literature, with only 7\% of reviewed articles addressing them.

Although attributes such as IP addresses and MAC addresses may not always be considered direct PII, they are classified as such under regulations such as GDPR due to their potential for re-identification when combined with other data. The presence of multiple attributes in logs increases this risk, necessitating a thorough evaluation of all exposed attributes for effective anonymization.

Based on our analyses of software log privacy from multiple perspectives, \textbf{we consider IP addresses, MAC addresses, hostnames, file paths, IDs, URLs, usernames, port numbers, and configuration details as generally sensitive information in software logs.} These attributes, particularly when combined, can disclose private user information or sensitive organizational details. Hence, a comprehensive anonymization strategy should consider all potential attribute combinations to prevent unintended disclosures.

\subsection{Research Gaps and Future Directions}
\begin{enumerate}[leftmargin=*]
\item \textit{Broadening the focus on diverse log attributes.} Comparing the findings from Tables~\ref{tab:attributes_preliminary},~\ref{tab:attributes_papers}, and~\ref{tab:attributes_industry}, it is clear that research papers often do not adequately cover a broad range of sensitive log attributes. As discussed earlier, while IP addresses are extensively studied in the literature, other attributes, such as MAC addresses or file paths, have not been sufficiently explored. Future research should address these gaps for a more comprehensive understanding of log privacy. Additionally, a significant portion of existing studies focuses on anonymizing timestamps to prevent attacks. However, timestamps are often essential for log utility, especially for key downstream tasks such as anomaly detection and performance prediction~\cite{wang2015performance, fronza2013failure}. We recommend future research focus on anonymizing other sensitive attributes that may reveal private information instead of relying heavily on timestamp modifications to better balance privacy and utility.

\item \textit{Developing specialized anonymization tools for software logs.} The unique characteristics of software logs (such as their unstructured nature, the combination of strings and numerical data, and the high volume of entries) requires the development of specialized anonymization tools customized for these logs. While there exist some log anonymization tools, they are either very specific to some log types (e.g., network logs), handle a limited number of sensitive attributes (e.g., only IP addresses), or require a structured input format, making them unsuitable for unstructured logs. Future efforts should focus on creating software log anonymization tools that can manage a broad range of log attributes and overcome challenges such as computational efficiency and processing time, enhancing both privacy protection and data utility.

\item \textit{Developing a privacy score for software logs.} As highlighted in this study, a combination of various log attributes can lead to significant privacy risks. Future research could focus on designing a comprehensive privacy scoring system for software logs. This privacy score would evaluate the sensitivity of a log line or log file by analyzing the presence and combination of specific attributes, such as IP addresses, user identifiers, and network-related information. By establishing a threshold based on this score, organizations could more effectively assess whether sharing a particular log would pose a privacy risk. Such a scoring system could serve as a practical tool for data controllers to automate privacy assessments, enforce privacy policies, and mitigate potential data breaches while still enabling necessary data analysis.

\item \textit{Examining privacy challenges in small-sized companies vs. larger organizations.} Future research could explore whether data privacy practitioners in small and medium-sized enterprises encounter unique challenges compared to those in larger organizations. Small companies often have fewer resources, less regulatory oversight, and limited expertise in log data privacy, which may impact their ability to implement effective anonymization and compliance measures. Investigating these differences could help tailor privacy solutions and regulatory frameworks to better support organizations of varying sizes.

\item \textit{Comparing stated privacy policies with actual log data collection.} Future research could investigate the gap between what companies claim in their privacy policies and the log data they actually collect. Analyzing discrepancies between stated commitments and real-world practices could provide insights into transparency issues and help develop stronger regulatory guidelines for log privacy.

\end{enumerate}

\section{Threats to Validity}
\label{sec:threats}

\textbf{Internal Validity.} 
The definition of ``sensitivity'' can differ between individuals and across industries. To mitigate this issue, we explored log privacy from multiple perspectives, including regulatory frameworks, academic research, and industry experts. 

\noindent \textbf{External Validity.} 
We analyzed 25 publicly available log datasets; however, the findings may not be applicable to other types of logs, such as those used in enterprise or government environments. Similarly, while our survey provides valuable insights, it captures the views of a specific group of professionals and may not fully represent all industries or global practices. To mitigate this risk, we included a relatively large sample size of 45 participants to capture a diverse range of opinions. Moreover, our study focuses on specific data privacy regulations, which could limit its relevance to regions with different legal frameworks. Nevertheless, we focused on the most widely recognized privacy regulations to ensure broader applicability.

\noindent \textbf{Construct Validity.} 
The survey questions may not capture all aspects of privacy concerns in software logs and could concentrate on certain log types or attributes. To mitigate this, we first conducted a preliminary study to identify the most common attributes in software logs and designed our survey accordingly. Additionally, the parsing tool used in our log analysis might have accuracy limitations, and some log attributes may have been missed. However, by combining insights from log analysis, data regulations, research articles, and industry experts, we aimed to achieve a well-rounded view of sensitive attributes in software logs.
\section{Conclusion}
\label{sec:conclusion}

We conduct an in-depth analysis of privacy in software logs by examining the issue from multiple perspectives: regulatory frameworks, research literature, and industry practices. Our analysis of 25 publicly available log datasets identifies the most common log attributes that may require anonymization, such as IP addresses, MAC addresses, and hostnames. Our review of privacy regulations, including GDPR and CCPA, highlights the legal obligations to protect such sensitive data. Additionally, our systematic literature review reveals that while IP addresses and timestamps are commonly studied, other attributes, such as MAC addresses and file paths, are underexplored. Insights from a survey of 45 industry professionals further emphasize the challenges in balancing privacy and utility in log data, as well as in dealing with computational costs and processing time.

By bridging gaps between regulatory requirements, academic research, and industry practices, our study lays the groundwork for a more unified approach to privacy in software logs. Our contributions include a clear definition of sensitive information in software logs, identifying key gaps in current research and industry practices, and emphasizing the importance of context when determining which log attributes should be anonymized. Future work should aim to broaden the focus on diverse log attributes and develop anonymization tools specific to software logs. Additionally, there is an opportunity to develop a privacy score that could evaluate the sensitivity of log entries based on their attributes and determine if sharing them would pose a privacy risk.
\section{Data Availability}
\label{sec:data_availability}

To support future replication or extension of our research, we have made our replication package available. This includes the list of log datasets analyzed, the list of reviewed articles, detailed survey questionnaire, and the survey responses. The package can be accessed at \url{https://github.com/mooselab/log_privacy}.
\section*{Acknowledgments}

We would like to gratefully acknowledge the Natural Sciences and Engineering Research Council of Canada (NSERC, \#RGPIN-2021-03900) and the Fonds de recherche du Québec – Nature et technologies (FRQNT, \#326866) for their funding support for this work.

\bibliographystyle{ACM-Reference-Format}
\bibliography{acmart}


\begin{thebibliography}{49}


\ifx \showCODEN    \undefined \def \showCODEN     #1{\unskip}     \fi
\ifx \showDOI      \undefined \def \showDOI       #1{#1}\fi
\ifx \showISBNx    \undefined \def \showISBNx     #1{\unskip}     \fi
\ifx \showISBNxiii \undefined \def \showISBNxiii  #1{\unskip}     \fi
\ifx \showISSN     \undefined \def \showISSN      #1{\unskip}     \fi
\ifx \showLCCN     \undefined \def \showLCCN      #1{\unskip}     \fi
\ifx \shownote     \undefined \def \shownote      #1{#1}          \fi
\ifx \showarticletitle \undefined \def \showarticletitle #1{#1}   \fi
\ifx \showURL      \undefined \def \showURL       {\relax}        \fi
\providecommand\bibfield[2]{#2}
\providecommand\bibinfo[2]{#2}
\providecommand\natexlab[1]{#1}
\providecommand\showeprint[2][]{arXiv:#2}

\bibitem[Aghili et~al\mbox{.}(2023)]%
        {aghili2023studying}
\bibfield{author}{\bibinfo{person}{Roozbeh Aghili}, \bibinfo{person}{Heng Li}, {and} \bibinfo{person}{Foutse Khomh}.} \bibinfo{year}{2023}\natexlab{}.
\newblock \showarticletitle{Studying the characteristics of AIOps projects on GitHub}.
\newblock \bibinfo{journal}{\emph{Empirical Software Engineering}} \bibinfo{volume}{28}, \bibinfo{number}{6} (\bibinfo{year}{2023}), \bibinfo{pages}{143}.
\newblock


\bibitem[Agrawal et~al\mbox{.}(2019)]%
        {agrawal2019delog}
\bibfield{author}{\bibinfo{person}{Amey Agrawal}, \bibinfo{person}{Abhishek Dixit}, \bibinfo{person}{Namrata~A Shettar}, \bibinfo{person}{Darshil Kapadia}, \bibinfo{person}{Vikram Agrawal}, \bibinfo{person}{Rajat Gupta}, {and} \bibinfo{person}{Rohit Karlupia}.} \bibinfo{year}{2019}\natexlab{}.
\newblock \showarticletitle{Delog: A high-performance privacy preserving log filtering framework}. In \bibinfo{booktitle}{\emph{2019 IEEE International Conference on Big Data (Big Data)}}. IEEE, \bibinfo{pages}{1739--1748}.
\newblock


\bibitem[Barbaro(2006)]%
        {aol}
\bibfield{author}{\bibinfo{person}{Tom Barbaro, Michael; Zeller~Jr}.} \bibinfo{year}{2006}\natexlab{}.
\newblock \bibinfo{title}{A Face Is Exposed for AOL Searcher No. 4417749}.
\newblock
\newblock
\urldef\tempurl%
\url{https://www.nytimes.com/2006/08/09/technology/09aol.html}
\showURL{%
\tempurl}
\newblock
\shownote{Accessed on September 11, 2024}.


\bibitem[Barik et~al\mbox{.}(2016)]%
        {barik2016bones}
\bibfield{author}{\bibinfo{person}{Titus Barik}, \bibinfo{person}{Robert DeLine}, \bibinfo{person}{Steven Drucker}, {and} \bibinfo{person}{Danyel Fisher}.} \bibinfo{year}{2016}\natexlab{}.
\newblock \showarticletitle{The bones of the system: A case study of logging and telemetry at microsoft}. In \bibinfo{booktitle}{\emph{Proceedings of the 38th International Conference on Software Engineering Companion}}. \bibinfo{pages}{92--101}.
\newblock


\bibitem[Batoun et~al\mbox{.}(2024)]%
        {batoun2024literature}
\bibfield{author}{\bibinfo{person}{Mohamed~Amine Batoun}, \bibinfo{person}{Mohammed Sayagh}, \bibinfo{person}{Roozbeh Aghili}, \bibinfo{person}{Ali Ouni}, {and} \bibinfo{person}{Heng Li}.} \bibinfo{year}{2024}\natexlab{}.
\newblock \showarticletitle{A literature review and existing challenges on software logging practices: From the creation to the analysis of software logs}.
\newblock \bibinfo{journal}{\emph{Empirical Software Engineering}} \bibinfo{volume}{29}, \bibinfo{number}{4} (\bibinfo{year}{2024}), \bibinfo{pages}{103}.
\newblock


\bibitem[Bogatinovski et~al\mbox{.}(2021)]%
        {bogatinovski2021artificial}
\bibfield{author}{\bibinfo{person}{Jasmin Bogatinovski}, \bibinfo{person}{Sasho Nedelkoski}, \bibinfo{person}{Alexander Acker}, \bibinfo{person}{Florian Schmidt}, \bibinfo{person}{Thorsten Wittkopp}, \bibinfo{person}{Soeren Becker}, \bibinfo{person}{Jorge Cardoso}, {and} \bibinfo{person}{Odej Kao}.} \bibinfo{year}{2021}\natexlab{}.
\newblock \showarticletitle{Artificial intelligence for it operations (aiops) workshop white paper}.
\newblock \bibinfo{journal}{\emph{arXiv preprint arXiv:2101.06054}} (\bibinfo{year}{2021}).
\newblock


\bibitem[Brekne and {\AA}rnes(2005)]%
        {brekne2005circumventing}
\bibfield{author}{\bibinfo{person}{T{\o}nnes Brekne} {and} \bibinfo{person}{Andr{\'e} {\AA}rnes}.} \bibinfo{year}{2005}\natexlab{}.
\newblock \showarticletitle{Circumventing IP-address pseudonymization.}. In \bibinfo{booktitle}{\emph{Communications and Computer Networks}}. \bibinfo{pages}{43--48}.
\newblock


\bibitem[Brekne et~al\mbox{.}(2005)]%
        {brekne2005anonymization}
\bibfield{author}{\bibinfo{person}{T{\o}nnes Brekne}, \bibinfo{person}{Andr{\'e} {\AA}rnes}, {and} \bibinfo{person}{Arne {\O}sleb{\o}}.} \bibinfo{year}{2005}\natexlab{}.
\newblock \showarticletitle{Anonymization of ip traffic monitoring data: Attacks on two prefix-preserving anonymization schemes and some proposed remedies}. In \bibinfo{booktitle}{\emph{International Workshop on Privacy Enhancing Technologies}}. Springer, \bibinfo{pages}{179--196}.
\newblock


\bibitem[Burkhart et~al\mbox{.}(2010)]%
        {burkhart2010role}
\bibfield{author}{\bibinfo{person}{Martin Burkhart}, \bibinfo{person}{Dominik Schatzmann}, \bibinfo{person}{Brian Trammell}, \bibinfo{person}{Elisa Boschi}, {and} \bibinfo{person}{Bernhard Plattner}.} \bibinfo{year}{2010}\natexlab{}.
\newblock \showarticletitle{The role of network trace anonymization under attack}.
\newblock \bibinfo{journal}{\emph{ACM SIGCOMM Computer Communication Review}} \bibinfo{volume}{40}, \bibinfo{number}{1} (\bibinfo{year}{2010}), \bibinfo{pages}{5--11}.
\newblock


\bibitem[Cinque et~al\mbox{.}(2012)]%
        {cinque2012event}
\bibfield{author}{\bibinfo{person}{Marcello Cinque}, \bibinfo{person}{Domenico Cotroneo}, {and} \bibinfo{person}{Antonio Pecchia}.} \bibinfo{year}{2012}\natexlab{}.
\newblock \showarticletitle{Event logs for the analysis of software failures: A rule-based approach}.
\newblock \bibinfo{journal}{\emph{IEEE Transactions on Software Engineering}} \bibinfo{volume}{39}, \bibinfo{number}{6} (\bibinfo{year}{2012}), \bibinfo{pages}{806--821}.
\newblock


\bibitem[Cooper(2008)]%
        {cooper2008survey}
\bibfield{author}{\bibinfo{person}{Alissa Cooper}.} \bibinfo{year}{2008}\natexlab{}.
\newblock \showarticletitle{A survey of query log privacy-enhancing techniques from a policy perspective}.
\newblock \bibinfo{journal}{\emph{ACM Transactions on the Web (TWEB)}} \bibinfo{volume}{2}, \bibinfo{number}{4} (\bibinfo{year}{2008}), \bibinfo{pages}{1--27}.
\newblock


\bibitem[for Standardization(2022)]%
        {iso27001}
\bibfield{author}{\bibinfo{person}{International~Organization for Standardization}.} \bibinfo{year}{2022}\natexlab{}.
\newblock \bibinfo{title}{ISO/IEC 27001:2022}.
\newblock
\newblock
\urldef\tempurl%
\url{https://www.iso.org/standard/27001}
\showURL{%
\tempurl}
\newblock
\shownote{Accessed on September 11, 2024}.


\bibitem[Foukarakis et~al\mbox{.}(2007)]%
        {foukarakis2007flexible}
\bibfield{author}{\bibinfo{person}{Michalis Foukarakis}, \bibinfo{person}{Demetres Antoniades}, \bibinfo{person}{Spiros Antonatos}, {and} \bibinfo{person}{Evangelos~P Markatos}.} \bibinfo{year}{2007}\natexlab{}.
\newblock \showarticletitle{Flexible and high-performance anonymization of NetFlow records using anontool}. In \bibinfo{booktitle}{\emph{2007 Third International Conference on Security and Privacy in Communications Networks and the Workshops-SecureComm 2007}}. IEEE, \bibinfo{pages}{33--38}.
\newblock


\bibitem[Foukarakis et~al\mbox{.}(2009)]%
        {foukarakis2009deep}
\bibfield{author}{\bibinfo{person}{Michael Foukarakis}, \bibinfo{person}{Demetres Antoniades}, {and} \bibinfo{person}{Michalis Polychronakis}.} \bibinfo{year}{2009}\natexlab{}.
\newblock \showarticletitle{Deep packet anonymization}. In \bibinfo{booktitle}{\emph{Proceedings of the Second European Workshop on System Security}}. \bibinfo{pages}{16--21}.
\newblock


\bibitem[Fronza et~al\mbox{.}(2013)]%
        {fronza2013failure}
\bibfield{author}{\bibinfo{person}{Ilenia Fronza}, \bibinfo{person}{Alberto Sillitti}, \bibinfo{person}{Giancarlo Succi}, \bibinfo{person}{Mikko Terho}, {and} \bibinfo{person}{Jelena Vlasenko}.} \bibinfo{year}{2013}\natexlab{}.
\newblock \showarticletitle{Failure prediction based on log files using random indexing and support vector machines}.
\newblock \bibinfo{journal}{\emph{Journal of Systems and Software}} \bibinfo{volume}{86}, \bibinfo{number}{1} (\bibinfo{year}{2013}), \bibinfo{pages}{2--11}.
\newblock


\bibitem[GDPRhub(2023)]%
        {breyer}
\bibfield{author}{\bibinfo{person}{GDPRhub}.} \bibinfo{year}{2023}\natexlab{}.
\newblock \bibinfo{title}{CJEU - C-582/14 - Patrick Breyer}.
\newblock
\newblock
\urldef\tempurl%
\url{https://gdprhub.eu/index.php?title=CJEU_-_C-582/14_-_Patrick_Breyer}
\showURL{%
\tempurl}
\newblock
\shownote{Accessed on September 11, 2024}.


\bibitem[Gu and Dong(2023)]%
        {gu2023pd}
\bibfield{author}{\bibinfo{person}{Xiaodan Gu} {and} \bibinfo{person}{Kai Dong}.} \bibinfo{year}{2023}\natexlab{}.
\newblock \showarticletitle{PD-PAn: Prefix-and Distribution-Preserving Internet of Things Traffic Anonymization}.
\newblock \bibinfo{journal}{\emph{Electronics}} \bibinfo{volume}{12}, \bibinfo{number}{20} (\bibinfo{year}{2023}), \bibinfo{pages}{4369}.
\newblock


\bibitem[Han et~al\mbox{.}(2020)]%
        {han2020aft}
\bibfield{author}{\bibinfo{person}{Chunjing Han}, \bibinfo{person}{Kunkun Sun}, \bibinfo{person}{Haina Tang}, \bibinfo{person}{Yulei Wu}, {and} \bibinfo{person}{Xiaodan Zhang}.} \bibinfo{year}{2020}\natexlab{}.
\newblock \showarticletitle{AFT-Anon: A scaling method for online trace anonymization based on anonymous flow tables}. In \bibinfo{booktitle}{\emph{2020 IEEE Symposium on Computers and Communications (ISCC)}}. IEEE, \bibinfo{pages}{1--7}.
\newblock


\bibitem[He et~al\mbox{.}(2017)]%
        {he2017drain}
\bibfield{author}{\bibinfo{person}{Pinjia He}, \bibinfo{person}{Jieming Zhu}, \bibinfo{person}{Zibin Zheng}, {and} \bibinfo{person}{Michael~R Lyu}.} \bibinfo{year}{2017}\natexlab{}.
\newblock \showarticletitle{Drain: An online log parsing approach with fixed depth tree}. In \bibinfo{booktitle}{\emph{2017 IEEE international conference on web services (ICWS)}}. IEEE, \bibinfo{pages}{33--40}.
\newblock


\bibitem[Jain et~al\mbox{.}(2016)]%
        {jain2016big}
\bibfield{author}{\bibinfo{person}{Priyank Jain}, \bibinfo{person}{Manasi Gyanchandani}, {and} \bibinfo{person}{Nilay Khare}.} \bibinfo{year}{2016}\natexlab{}.
\newblock \showarticletitle{Big data privacy: a technological perspective and review}.
\newblock \bibinfo{journal}{\emph{Journal of Big Data}}  \bibinfo{volume}{3} (\bibinfo{year}{2016}), \bibinfo{pages}{1--25}.
\newblock


\bibitem[King et~al\mbox{.}(2009)]%
        {king2009taxonomy}
\bibfield{author}{\bibinfo{person}{Justin King}, \bibinfo{person}{Kiran Lakkaraju}, {and} \bibinfo{person}{Adam Slagell}.} \bibinfo{year}{2009}\natexlab{}.
\newblock \showarticletitle{A taxonomy and adversarial model for attacks against network log anonymization}. In \bibinfo{booktitle}{\emph{Proceedings of the 2009 ACM symposium on Applied Computing}}. \bibinfo{pages}{1286--1293}.
\newblock


\bibitem[Kitchenham et~al\mbox{.}(2015)]%
        {kitchenham2015evidence}
\bibfield{author}{\bibinfo{person}{Barbara~Ann Kitchenham}, \bibinfo{person}{David Budgen}, {and} \bibinfo{person}{Pearl Brereton}.} \bibinfo{year}{2015}\natexlab{}.
\newblock \bibinfo{booktitle}{\emph{Evidence-based software engineering and systematic reviews}}. Vol.~\bibinfo{volume}{4}.
\newblock \bibinfo{publisher}{CRC press}.
\newblock


\bibitem[Kitchenham and Pfleeger(2008)]%
        {kitchenham2008personal}
\bibfield{author}{\bibinfo{person}{Barbara~A Kitchenham} {and} \bibinfo{person}{Shari~L Pfleeger}.} \bibinfo{year}{2008}\natexlab{}.
\newblock \showarticletitle{Personal opinion surveys}.
\newblock In \bibinfo{booktitle}{\emph{Guide to advanced empirical software engineering}}. \bibinfo{publisher}{Springer}, \bibinfo{pages}{63--92}.
\newblock


\bibitem[Laboratory(1998)]%
        {ita}
\bibfield{author}{\bibinfo{person}{Lawrence Berkeley~National Laboratory}.} \bibinfo{year}{1998}\natexlab{}.
\newblock \bibinfo{title}{The Internet Traffic Archive}.
\newblock
\newblock
\urldef\tempurl%
\url{https://ita.ee.lbl.gov/html/traces.html}
\showURL{%
\tempurl}
\newblock
\shownote{Accessed on September 11, 2024}.


\bibitem[Li et~al\mbox{.}(2020)]%
        {li2020qualitative}
\bibfield{author}{\bibinfo{person}{Heng Li}, \bibinfo{person}{Weiyi Shang}, \bibinfo{person}{Bram Adams}, \bibinfo{person}{Mohammed Sayagh}, {and} \bibinfo{person}{Ahmed~E Hassan}.} \bibinfo{year}{2020}\natexlab{}.
\newblock \showarticletitle{A qualitative study of the benefits and costs of logging from developers’ perspectives}.
\newblock \bibinfo{journal}{\emph{IEEE Transactions on Software Engineering}} \bibinfo{volume}{47}, \bibinfo{number}{12} (\bibinfo{year}{2020}), \bibinfo{pages}{2858--2873}.
\newblock


\bibitem[Li et~al\mbox{.}(2023)]%
        {li2023triplelp}
\bibfield{author}{\bibinfo{person}{Teng Li}, \bibinfo{person}{Shengkai Zhang}, \bibinfo{person}{Zexu Dang}, \bibinfo{person}{Yongcai Xiao}, {and} \bibinfo{person}{Zhuo Ma}.} \bibinfo{year}{2023}\natexlab{}.
\newblock \showarticletitle{TripleLP: Privacy-Preserving Log Parsing Based on Blockchain}. In \bibinfo{booktitle}{\emph{2023 IEEE 14th International Symposium on Parallel Architectures, Algorithms and Programming (PAAP)}}. IEEE, \bibinfo{pages}{1--6}.
\newblock


\bibitem[Li et~al\mbox{.}(2005)]%
        {li2005canine}
\bibfield{author}{\bibinfo{person}{Yifan Li}, \bibinfo{person}{Adam Slagell}, \bibinfo{person}{Katherine Luo}, {and} \bibinfo{person}{William Yurcik}.} \bibinfo{year}{2005}\natexlab{}.
\newblock \showarticletitle{Canine: A combined conversion and anonymization tool for processing netflows for security}. In \bibinfo{booktitle}{\emph{International conference on telecommunication systems modeling and analysis}}, Vol.~\bibinfo{volume}{21}. Citeseer.
\newblock


\bibitem[Majeed et~al\mbox{.}(2022)]%
        {majeed2022toward}
\bibfield{author}{\bibinfo{person}{Abdul Majeed}, \bibinfo{person}{Safiullah Khan}, {and} \bibinfo{person}{Seong~Oun Hwang}.} \bibinfo{year}{2022}\natexlab{}.
\newblock \showarticletitle{Toward privacy preservation using clustering based anonymization: recent advances and future research outlook}.
\newblock \bibinfo{journal}{\emph{IEEE Access}}  \bibinfo{volume}{10} (\bibinfo{year}{2022}), \bibinfo{pages}{53066--53097}.
\newblock


\bibitem[Manocchio et~al\mbox{.}(2024)]%
        {manocchio2024configurable}
\bibfield{author}{\bibinfo{person}{Liam~Daly Manocchio}, \bibinfo{person}{Siamak Layeghy}, \bibinfo{person}{David Gwynne}, {and} \bibinfo{person}{Marius Portmann}.} \bibinfo{year}{2024}\natexlab{}.
\newblock \showarticletitle{A configurable anonymisation approach for network flow data: Balancing utility and privacy}.
\newblock \bibinfo{journal}{\emph{Computers and Electrical Engineering}}  \bibinfo{volume}{118} (\bibinfo{year}{2024}), \bibinfo{pages}{109465}.
\newblock


\bibitem[McSherry and Mahajan(2010)]%
        {mcsherry2010differentially}
\bibfield{author}{\bibinfo{person}{Frank McSherry} {and} \bibinfo{person}{Ratul Mahajan}.} \bibinfo{year}{2010}\natexlab{}.
\newblock \showarticletitle{Differentially-private network trace analysis}.
\newblock \bibinfo{journal}{\emph{ACM SIGCOMM Computer Communication Review}} \bibinfo{volume}{40}, \bibinfo{number}{4} (\bibinfo{year}{2010}), \bibinfo{pages}{123--134}.
\newblock


\bibitem[Michael~Zink and Kurose(2008)]%
        {youtube}
\bibfield{author}{\bibinfo{person}{Yu~Gu Michael~Zink, Kyoungwon~Suh} {and} \bibinfo{person}{Jim Kurose}.} \bibinfo{year}{2008}\natexlab{}.
\newblock \bibinfo{title}{YouTube Traces From the Campus Network}.
\newblock
\newblock
\urldef\tempurl%
\url{https://traces.cs.umass.edu/index.php/Network/Network}
\showURL{%
\tempurl}
\newblock
\shownote{Accessed on September 11, 2024}.


\bibitem[Minshall(2005)]%
        {tcpdpriv}
\bibfield{author}{\bibinfo{person}{Greg Minshall}.} \bibinfo{year}{2005}\natexlab{}.
\newblock \bibinfo{title}{TCPDPRIV}.
\newblock
\newblock
\urldef\tempurl%
\url{https://ita.ee.lbl.gov/html/contrib/tcpdpriv.html}
\showURL{%
\tempurl}
\newblock
\shownote{Accessed on September 11, 2024}.


\bibitem[Mohammady et~al\mbox{.}(2018)]%
        {mohammady2018preserving}
\bibfield{author}{\bibinfo{person}{Meisam Mohammady}, \bibinfo{person}{Lingyu Wang}, \bibinfo{person}{Yuan Hong}, \bibinfo{person}{Habib Louafi}, \bibinfo{person}{Makan Pourzandi}, {and} \bibinfo{person}{Mourad Debbabi}.} \bibinfo{year}{2018}\natexlab{}.
\newblock \showarticletitle{Preserving both privacy and utility in network trace anonymization}. In \bibinfo{booktitle}{\emph{Proceedings of the 2018 ACM SIGSAC Conference on Computer and Communications Security}}. \bibinfo{pages}{459--474}.
\newblock


\bibitem[Narayanan and Shmatikov(2006)]%
        {narayanan2006break}
\bibfield{author}{\bibinfo{person}{Arvind Narayanan} {and} \bibinfo{person}{Vitaly Shmatikov}.} \bibinfo{year}{2006}\natexlab{}.
\newblock \showarticletitle{How to break anonymity of the netflix prize dataset}.
\newblock \bibinfo{journal}{\emph{arXiv preprint cs/0610105}} (\bibinfo{year}{2006}).
\newblock


\bibitem[of~California(2018)]%
        {ccpa}
\bibfield{author}{\bibinfo{person}{U.S.~State of California}.} \bibinfo{year}{2018}\natexlab{}.
\newblock \bibinfo{title}{California Consumer Privacy Act}.
\newblock
\newblock
\urldef\tempurl%
\url{https://oag.ca.gov/privacy/ccpa}
\showURL{%
\tempurl}
\newblock
\shownote{Accessed on September 11, 2024}.


\bibitem[of~Canada(2000)]%
        {pipeda}
\bibfield{author}{\bibinfo{person}{Privacy~Commissioner of Canada}.} \bibinfo{year}{2000}\natexlab{}.
\newblock \bibinfo{title}{Personal Information Protection and Electronic Documents Act}.
\newblock
\newblock
\urldef\tempurl%
\url{https://www.priv.gc.ca/en/privacy-topics/privacy-laws-in-canada/the-personal-information-protection-and-electronic-documents-act-pipeda/}
\showURL{%
\tempurl}
\newblock
\shownote{Accessed on September 11, 2024}.


\bibitem[of~Health and Services(1996)]%
        {hipaa}
\bibfield{author}{\bibinfo{person}{U.S.~Department of Health} {and} \bibinfo{person}{Human Services}.} \bibinfo{year}{1996}\natexlab{}.
\newblock \bibinfo{title}{Health Insurance Portability and Accountability Act}.
\newblock
\newblock
\urldef\tempurl%
\url{https://www.hhs.gov/hipaa/index.html}
\showURL{%
\tempurl}
\newblock
\shownote{Accessed on September 11, 2024}.


\bibitem[Plonka(2003)]%
        {ip2anonip}
\bibfield{author}{\bibinfo{person}{Dave Plonka}.} \bibinfo{year}{2003}\natexlab{}.
\newblock \bibinfo{title}{ip2anonip}.
\newblock
\newblock
\urldef\tempurl%
\url{https://pages.cs.wisc.edu/%7Eplonka/ip2anonip/}
\showURL{%
\tempurl}
\newblock
\shownote{Accessed on September 11, 2024}.


\bibitem[Qin et~al\mbox{.}(2024)]%
        {qin2024preprocessing}
\bibfield{author}{\bibinfo{person}{Qiaolin Qin}, \bibinfo{person}{Roozbeh Aghili}, \bibinfo{person}{Heng Li}, {and} \bibinfo{person}{Ettore Merlo}.} \bibinfo{year}{2024}\natexlab{}.
\newblock \showarticletitle{Preprocessing is All You Need: Boosting the Performance of Log Parsers With a General Preprocessing Framework}.
\newblock \bibinfo{journal}{\emph{arXiv preprint arXiv:2412.05254}} (\bibinfo{year}{2024}).
\newblock


\bibitem[Silva et~al\mbox{.}(2021)]%
        {silva2021privacy}
\bibfield{author}{\bibinfo{person}{Paulo Silva}, \bibinfo{person}{Edmundo Monteiro}, {and} \bibinfo{person}{Paulo Simoes}.} \bibinfo{year}{2021}\natexlab{}.
\newblock \showarticletitle{Privacy in the cloud: A survey of existing solutions and research challenges}.
\newblock \bibinfo{journal}{\emph{IEEE access}}  \bibinfo{volume}{9} (\bibinfo{year}{2021}), \bibinfo{pages}{10473--10497}.
\newblock


\bibitem[Slagell and Yurcik(2005)]%
        {slagell2005sharing}
\bibfield{author}{\bibinfo{person}{Adam Slagell} {and} \bibinfo{person}{William Yurcik}.} \bibinfo{year}{2005}\natexlab{}.
\newblock \showarticletitle{Sharing computer network logs for security and privacy: A motivation for new methodologies of anonymization}. In \bibinfo{booktitle}{\emph{Workshop of the 1st International Conference on Security and Privacy for Emerging Areas in Communication Networks, 2005.}} IEEE, \bibinfo{pages}{80--89}.
\newblock


\bibitem[Slagell et~al\mbox{.}(2006)]%
        {slagell2006flaim}
\bibfield{author}{\bibinfo{person}{Adam~J Slagell}, \bibinfo{person}{Kiran Lakkaraju}, {and} \bibinfo{person}{Katherine Luo}.} \bibinfo{year}{2006}\natexlab{}.
\newblock \showarticletitle{FLAIM: A Multi-level Anonymization Framework for Computer and Network Logs.}. In \bibinfo{booktitle}{\emph{LISA}}, Vol.~\bibinfo{volume}{6}. \bibinfo{pages}{3--8}.
\newblock


\bibitem[Sweeney(2002)]%
        {sweeney2002k}
\bibfield{author}{\bibinfo{person}{Latanya Sweeney}.} \bibinfo{year}{2002}\natexlab{}.
\newblock \showarticletitle{k-anonymity: A model for protecting privacy}.
\newblock \bibinfo{journal}{\emph{International journal of uncertainty, fuzziness and knowledge-based systems}} \bibinfo{volume}{10}, \bibinfo{number}{05} (\bibinfo{year}{2002}), \bibinfo{pages}{557--570}.
\newblock


\bibitem[Union(2022)]%
        {gdpr}
\bibfield{author}{\bibinfo{person}{European Union}.} \bibinfo{year}{2022}\natexlab{}.
\newblock \bibinfo{title}{General Data Protection Regulation}.
\newblock
\newblock
\urldef\tempurl%
\url{https://gdpr-info.eu/}
\showURL{%
\tempurl}
\newblock
\shownote{Accessed on September 11, 2024}.


\bibitem[Wang and Khan(2015)]%
        {wang2015performance}
\bibfield{author}{\bibinfo{person}{Kewen Wang} {and} \bibinfo{person}{Mohammad Maifi~Hasan Khan}.} \bibinfo{year}{2015}\natexlab{}.
\newblock \showarticletitle{Performance prediction for apache spark platform}. In \bibinfo{booktitle}{\emph{2015 IEEE 17th International Conference on High Performance Computing and Communications, 2015 IEEE 7th International Symposium on Cyberspace Safety and Security, and 2015 IEEE 12th International Conference on Embedded Software and Systems}}. IEEE, \bibinfo{pages}{166--173}.
\newblock


\bibitem[Wohlin et~al\mbox{.}(2012)]%
        {wohlin2012experimentation}
\bibfield{author}{\bibinfo{person}{Claes Wohlin}, \bibinfo{person}{Per Runeson}, \bibinfo{person}{Martin H{\"o}st}, \bibinfo{person}{Magnus~C Ohlsson}, \bibinfo{person}{Bj{\"o}rn Regnell}, {and} \bibinfo{person}{Anders Wessl{\'e}n}.} \bibinfo{year}{2012}\natexlab{}.
\newblock \bibinfo{booktitle}{\emph{Experimentation in software engineering}}.
\newblock \bibinfo{publisher}{Springer Science \& Business Media}.
\newblock


\bibitem[Xu et~al\mbox{.}(2002)]%
        {xu2002prefix}
\bibfield{author}{\bibinfo{person}{Jun Xu}, \bibinfo{person}{Jinliang Fan}, \bibinfo{person}{Mostafa~H Ammar}, {and} \bibinfo{person}{Sue~B Moon}.} \bibinfo{year}{2002}\natexlab{}.
\newblock \showarticletitle{Prefix-preserving ip address anonymization: Measurement-based security evaluation and a new cryptography-based scheme}. In \bibinfo{booktitle}{\emph{10th IEEE International Conference on Network Protocols, 2002. Proceedings.}} IEEE, \bibinfo{pages}{280--289}.
\newblock


\bibitem[Yuan et~al\mbox{.}(2010)]%
        {yuan2010sherlog}
\bibfield{author}{\bibinfo{person}{Ding Yuan}, \bibinfo{person}{Haohui Mai}, \bibinfo{person}{Weiwei Xiong}, \bibinfo{person}{Lin Tan}, \bibinfo{person}{Yuanyuan Zhou}, {and} \bibinfo{person}{Shankar Pasupathy}.} \bibinfo{year}{2010}\natexlab{}.
\newblock \showarticletitle{Sherlog: error diagnosis by connecting clues from run-time logs}. In \bibinfo{booktitle}{\emph{Proceedings of the fifteenth International Conference on Architectural support for programming languages and operating systems}}. \bibinfo{pages}{143--154}.
\newblock


\bibitem[Zhu et~al\mbox{.}(2023)]%
        {zhu2023loghub}
\bibfield{author}{\bibinfo{person}{Jieming Zhu}, \bibinfo{person}{Shilin He}, \bibinfo{person}{Pinjia He}, \bibinfo{person}{Jinyang Liu}, {and} \bibinfo{person}{Michael~R Lyu}.} \bibinfo{year}{2023}\natexlab{}.
\newblock \showarticletitle{Loghub: A large collection of system log datasets for ai-driven log analytics}. In \bibinfo{booktitle}{\emph{2023 IEEE 34th International Symposium on Software Reliability Engineering (ISSRE)}}. IEEE, \bibinfo{pages}{355--366}.
\newblock


\end{thebibliography}

\appendix

\end{document}